# Experimental Study of a Lorentz Actuated Orbit

**William R. Gorman and James D. Brownridge**

State University of New York at Binghamton
Department of Physics, Applied Physics, and Astronomy
Binghamton, NY 13902

**Mason Peck**

Cornell University
Sibley School of Mechanical and Aerospace Engineering
Ithaca, NY 14853

**This experimental study investigates a new technique to keep a satellite in orbit utilizing electrodynamics. The technique consists of establishing a charge on a satellite such that the body's motion through a planetary magnetic field induces acceleration via the Lorentz force. In order to find the relationship between capacitance and power required to balance incident plasma current, various objects were tested in high vacuum, plasma, and Xenon gas to determine their ability to hold charge. Radioactive material (Am-241) and pyroelectric crystals were tested as a candidate power source for charging the objects. Microscopic arcing was observed at voltages as low as -300 V. This arcing caused solder to explode off of the object. Insulating the object allowed the charge to remain on the object longer, while in the plasma, and also eliminated the arcing. However, this insulation does not allow a net charge to reside on the surface of the spacecraft.**

The driving mechanism behind the Lorentz Actuated Orbit is the well understood Lorentz force. This propulsion architecture was proposed by M. Peck and his team at Cornell University and differs from similar electromagnetic (EM) based propulsion systems because it is a propellantless propulsion system that does not require a tether or a solar sail.[1,2] The principle behind this idea is that the spacecraft itself would act as a charged body. This charged spacecraft would be traveling through planetary magnetic fields and would thus experience a force perpendicular to the planet's local, moving magnetic field. In general, some component of the force projects onto the satellite's inertial velocity, changing the satellite's orbital energy. With the proper design, the force could be controlled and therefore used as a means for propulsion. One of the key benefits of this approach is that it is propellantless, using only a source of charged particles to charge the craft. The two most promising types of particle sources proposed are Polonium-210 (Po-210) and pyroelectric crystals. However, due to recent pressures in obtaining Po-210, it was decided that Americium-241 (Am-241) would be used instead. This experimental study focused on creating test structures meant to remain charged in plasma for an extended period of time and on testing the performance of the particle sources in the plasma.

The experiments were conducted in a vacuum chamber equipped to be under high vacuum ($10^{-7}$ torr) and to contain plasman Xenon (Xe) gas. To create the plasma, a Busek Hollow Cathode plasma gun BHC-1500 was used in the form of its design as a neutralizer for Busek's 200

W Hall Thruster.[3] Charging the object in the chamber with it connected to one side of a capacitor allowed us to monitor the transfer of charge to the other side of the capacitor with the use of an ammeter, as shown in Fig. 1.

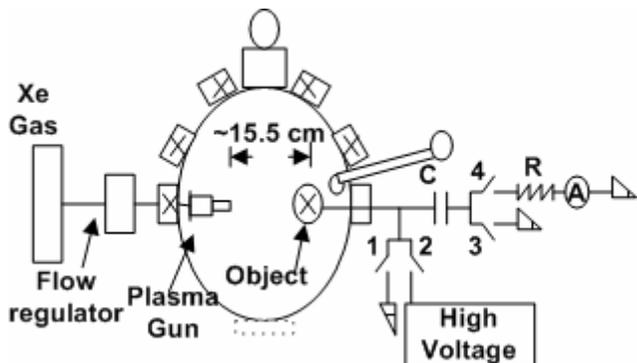

FIG. 1. The chamber setup for the experiments with the R-C circuit setup included. Switches 2 and 3 were linked and were opposite of 4. I.e., when 2 and 3 were closed, 4 was open. Switch 1 was always open, unless we were measuring the object connected to ground. The capacitor (C) was 0.005 μF and the resistor (R) was $1 \times 10^{12}$ Ω (10%).

The environment to which the object is subjected alters the rate at which the charge is transferred through the ammeter. Several solid and mesh materials were tested in four different environments: high vacuum, neutral Xe plasma, Xe gas, and with the object grounded. The materials used for the objects were copper, aluminum, tin and germanium, with the sizes of each object varying but in all cases smaller than a 6.4 cm radius, 13 cm high, right cylinder. Current for charging was provided by a high-voltage power supply at -1000V, except for the cases in which a 20 μCi Am-241 source was tested as a possible particle source.

A Langmuir probe[4-9] was used to characterize the plasma. The electron temperature ($T_e$) was found to be $2.54 \leq T_e \leq 4.43$ eV, with the electron density ($n_e$) being $4.03 \leq n_e \leq 5.32 \times 10^9$ cm$^{-3}$ depending upon the setting of the plasma gun. The decay of the -1000V charge was measured by an ammeter, as shown in Fig. 2. Since all of the objects, materials and size, behaved in essentially the same manner, only one is shown here.

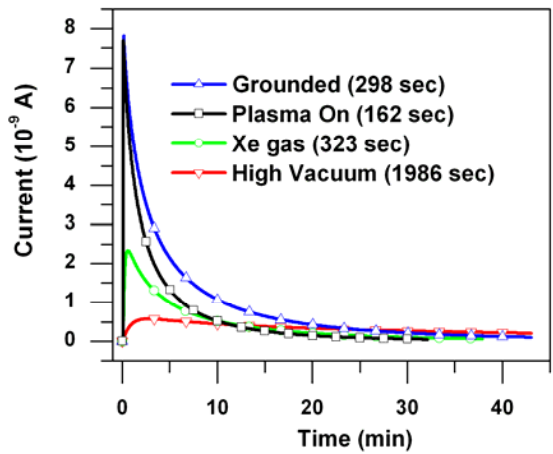

FIG. 2. Current versus time for the object in each environment. The numbers in the legend represent the time constant for the exponential decay of that particular curve.

While the -1000 V was being applied, arcing was observed on and around the object while it was in the plasma environment. This arcing visually appeared to be a single point explosion rather than a continuous static charge arc. After the arcing occurred, fragments of material were found in the chamber. X-ray spectroscopy confirmed it to be the solder ($Sn_{60}Pb_{40}$) that was used on the objects. By scanning the solder still remaining on the object, we were able to locate microscopic craters that represent strong evidence for the solder exploding off of the object, as seen in Fig. 3.

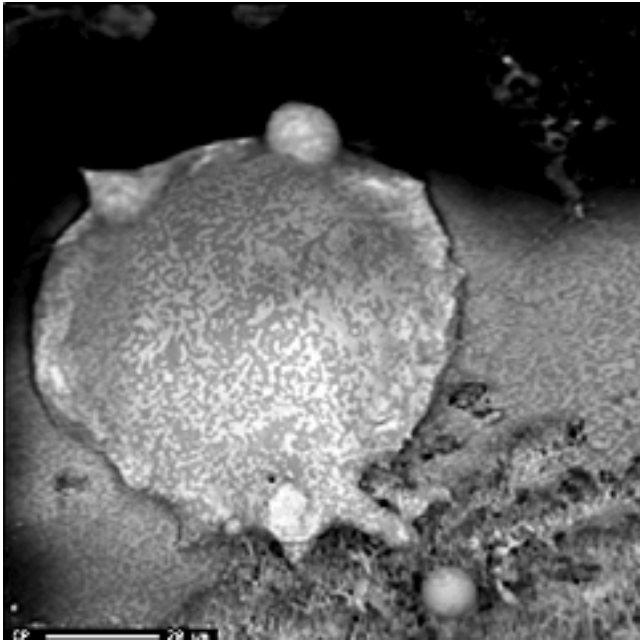

FIG. 3. Backscattered electron image of a crater in solder after -1000 V was applied to the object and it arced. This crater is believed to be responsible for the solder that was found in the chamber. The higher the atomic number of the element, the brighter the color in the image (the difference between the Sn and Pb can easily be seen inside the crater). The darker area near the top of the image is the edge of an oxide layer (easily seen in other images) which might be allowing differential charge to develop and be the cause of the arcing. The solid bar at the bottom left part of the image represents 20 μm.

We were able to observe the arcing with only -300 V applied, and the arcing subsided as the voltage was left on for an extended period of time. We were unable to apply +1000 V to the object because it would overload our power supply (10 mA maximum). However, we were able to melt Indium (157°C melting point) with only +300 V applied. The study of the Am-241 source used to charge an object was done in high vacuum ($10^{-7}$ torr). The results suggest that to charge an object to +1000 V, an 1170 μCi source would be needed.

The RC circuit setup shown in Fig. 1 helped amplify the effects seen and produced a different current curve depending upon the environment that the object was subjected to. In the case of a grounded object, as shown in Fig. 2, the current spiked and then quickly decayed, which is expected because the charge was being removed extremely fast. It took substantially longer for the charge to leave the object in the case of the high vacuum. For the case with the plasma on, the charge was removed from the object with a lower time constant than when it was connected to ground. This fast discharge impeded the progress of the experiment. To overcome this problem, we insulated the object by placing the object in a jar, pulled a vacuum in the jar, and sealed it. Then we placed the jar in the plasma. This effectively solved the problem of charge leaking into the plasma; however, an equal and opposite charge developed on the outside of the insulating material and caused the object to appear to have a net zero charge, preventing a Lorentz force.

Another issue is the problem of arcing, of common concern in spacecraft design, but which could not be reliably reproduced in the experimental setup. This arcing is believed to be caused by differential charging on the microscopic length scale. Several tangential experiments studied the arcing with a copper wire having different surface characteristics (scored, cleaned, oxidized, and oiled). These experiments could not point to a certain characteristic that caused it to occur regularly. The arcing did occur on both the solder and the bare metal, but only the solder was shown to become soft enough to explode off. The backscattered electron images of the arced solder show craters that exist either on an oxidized/non-oxidized boundary or in the middle of an oxidized layer. This arcing causes concern for any electrical equipment[10] but can be eliminated by simply insulating the solder/object.

The radioactive material and pyroelectric crystals were not tested in the plasma because arcing across these materials would potentially contaminate the chamber. Furthermore, the required current to charge an object in the plasma is much greater than anyone of the sources could produce.[11]

This experimental study of a Lorentz Actuated Orbit has provided considerable insight. The arcing that occurs at low voltages is a serious problem but can essentially be eliminated if the wires and solder are well insulated. The object can easily be charged in plasma only if it is insulated in such a way that it is not in direct contact with the plasma. The only problem then is the net zero charge that is observed by the magnetic field. The plasma itself, created in the lab, is not exactly identical to that of outer space or the upper atmosphere. Even though the electron temperature closely relates to those regions, the electron density is four orders of magnitude higher and may cause different results than one would obtain in a lower electron density environment.[5,10,12] Experimental research on a Lorentz Actuated Orbit is still ongoing with the hope that one day we will be able to establish an experimental basis for sizing a propellantless propulsion system.

ACKNOWLEDGEMENTS
The authors would like to thank B. Blackburn and Binghamton University's Physics, Applied Physics, and Astronomy department. This project was supported by NASA grant contract NAS5-03110, Subcontract Agreement No. 7605-003-071.

*Corresponding Author: William R. Gorman (wgorman1@binghamton.edu)